\documentstyle[12pt]{article}
\begin{document}
\begin{flushright}
arXiv: 0711.2389 [hep-th] \\ CAS-PHYS-BHU/Preprint
\end{flushright}
\vskip 2.5cm
\begin{center}
{\bf   SUPERFIELD APPROACH TO SYMMETRY INVARIANCE IN QED \\WITH COMPLEX SCALAR FIELDS}

\vskip 2.5cm

 R. P. MALIK\footnote{Presently also associated
with the DST Centre for Interdisciplinary Mathematical Sciences, Faculty of Science,
Banaras Hindu University, Varanasi - 221 005 (U. P.).} and B. P. MANDAL\\ {\it Department of Physics, 
Centre of Advanced Studies,}\\ {\it Banaras Hindu University, Varanasi-221 005, (U. P.),
India} \\  {\small   E-mails: malik@bhu.ac.in ; bhabani@bhu.ac.in }\\

\vskip 2cm

\end{center}

\noindent {\bf Abstract}: 
We show that the Grassmannian independence of the super Lagrangian density, expressed in terms of the superfields defined on a (4, 2)-dimensional supermanifold, is a clear-cut proof for the Becchi-Rouet-Stora-Tyutin (BRST) and anti-BRST invariance of the corresoponding four (3 + 1)-dimensional (4D) Lagrangian density that describes the interaction between the U(1) gauge field and the charged complex scalar fields. The above 4D field theoretical model is considered on a (4, 2)-dimensional supermanifold parametrized by the ordinary four spacetime variables $x^\mu$ (with $\mu = 0, 1, 2, 3$) and a pair of Grassmannian variables $\theta$ and $\bar\theta$ (with $\theta^2 = \bar\theta^2 = 0, \theta \bar\theta + \bar\theta \theta = 0$).
Geometrically, the (anti-)BRST invariance is encoded in the translation of the super Lagrangian density along the Grassmannian directions of the above supermanifold such that the outcome of this shift operation is zero.\\ 

\baselineskip=16pt



\newpage

\noindent
{\bf 1 Introduction}\\

\noindent
The usual superfield approach to BRST formalism [1-11] has been quite successful in providing the geometrical basis for the origin of the nilpotent (anti-)BRST symmetry transformations   and their corresponding nilpotent generators as far as the gauge and (anti-)ghost fields of the 4D (non-)Abelian 1-form gauge field theories are concerned. This approach has also been able to shed light on the geometrical meaning of the nilpotency and anticommutativity properties of the above transformations. It could not, however, explain the 
origin behind the
existence of the (anti-)BRST symmetry transformations that are associated with the matter fields of the above {\it interacting} 1-form gauge theories (see, e.g. [4-7]).

One of the key ingredients of the above superfield approach is the horizontality condition (HC)
which has been christened as the soul-flatness condition in [12]. In a set of research papers
[13-25], the above HC has been consistently extended (and, in some sense, generalized) so as to
derive the nilpotent (anti-)BRST symmetry transformations associated with the matter (e.g. Dirac, complex scalar, etc.) fields
within the framework of the superfield formulation. The salient features of the latter superfield approach \footnote{The consistent extensions 
(and, in some sense, generalizations)
of the usual superfield formulation have been collectively called by us as the augmented superfield approach to BRST formalism.} are as follows:

(i) In addition to the nilpotent (anti-)BRST symmetry transformations for the gauge and (anti-)ghost fields, the (anti-)BRST symmetry transformations for the matter fields are also derived precisely for a given interacting (non-)Abelian 1-form gauge theory, and

(ii) The geometrical interpretations of the (anti-)BRST symmetry transformations and their corresponding generators remain the same as the ones deduced due to the application of the HC {\it alone} in the context of the usual superfield approach to BRST formalism.

In
all the above attempts [1-25], the (anti-)BRST invariance of the 4D Lagrangian density has not yet been able to be captured within the framework of the superfield formulation. As a consequence, the geometrical basis for the existence of the (anti-)BRST invariance has been lacking for quite sometime within the framework of the above formulation [4-7]. It is, however,
gratifying to mention that,
in a very recent set of papers [26-28],
the above issue has been addressed for the (non-)Abelian 1-form gauge theories (with and without matter fields) and the geometrical origin 
for the existence of the (anti-)BRST invariance has been provided within the framework of the augmented superfield formulation.

To be precise, the subject matter of [26] concerns itself with the geometrical interpretations for the (anti-)BRST invariance of the (non-)Abelian 1-form gauge theories where there is {\it no} interaction with matter fields. The material of the works in [27,28] is connected with the above geometrical interpretation of the (anti-)BRST invariance for the Abelian and non-Abelian 1-form gauge theories where there is an explicit coupling between the 1-form gauge fields and the Dirac fields. It has been established, in the above works [26-28], that the (anti-)BRST invariance of the 4D Lagrangian densities is encoded in the Grassmannian 
independence of the super Lagrangian densities, defined in terms of the superfields, on an
appropriately chosen (4, 2)-dimensional supermanifold.

The purpose of our present paper is to capture the (anti-)BRST invariance of a new {\it interacting} field theoretical model within the framework of the augmented superfield formulation which is different from the models discussed in [26-28]. We demonstrate that,
like our earlier works [26-28], the (anti-)BRST invariance of the Lagrangian density of a 4D interacting U(1) Abelian gauge theory (where there is an explicit coupling between the gauge field and charged complex scalar fields) is encompassed in the proof that the corresponding super Lagrangian density, expressed in terms of certain specific (4, 2)-dimensional superfields, is independent of the Grassmannian variables of the  (4, 2)-dimensional supermanifold on which the above 4D theory is considered. In other words, we show that the operation of the partial derivatives
w.r.t. the Grassmannian variables, on the above (4, 2)-dimensional super Lagrangian density, is zero for the existence of the (anti-)BRST invariance.

The motivating factors behind the present investigation are as follows. First and foremost, it is important for us to check the validity of our proposal and observations, made in [26-28], for the new set of field theoretical models so that our basic ideas can be put on a firmer footing. Second, we have specifically chosen the field theoretic model of QED with complex scalar fields because it is a very {\it special} field theoretical model that allows the inclusion of the gauge (i.e. BRST) invariant quartic self-interacting term. Furthermore, for this model, the Noether conserved current (that couples with the U(1) gauge field to generate the interaction term) contains the gauge field itself. Finally, many interesting phenomena (e.g. Goldstone theorem, Higgs mechanism, spontaneous symmetry breaking, etc.) are associated with it which are {\it not} 
found in the case of QED with Dirac fields.

Our present paper is organized as follows. In section 2, we recapitulate the bare essentials of the nilpotent (anti-)BRST symmetry transformations in the Lagrangian formulation and express the gauge-fixing and Faddeev-Popov ghost terms of the (anti-)BRST invariant Lagrangian density of the 4D theory in {\it three} different forms. Section 3 is devoted to the discussion of the horizontality condition and its ramifications on the super Lagrangian density of the theory. Our section 4 deals with a gauge-invariant restriction on the matter superfields that enables us to express the 4D kinetic energy term of the matter (i.e. charged complex scalar) fields in terms of the matter superfields and the super covariant derivatives. In section 5, we derive the consequences of the previous sections 3 and 4 in one stroke by exploiting a gauge-invariant restriction on the matter superfields. Finally, we summarize our key results and point out future directions for further investigations in section 6.\\

\noindent
{\bf 2 (Anti-)BRST Symmetry Transformations:  A Brief Sketch}\\

\noindent
We begin with the (anti-)BRST invariant Lagrangian density of the 
4D\footnote{We adopt here the conventions and notations such that the 4D flat Minkowskian 
metric $\eta_{\mu\nu}$ is with signature $(+1, -1, -1, -1)$ and $(\partial \cdot A) = \partial_0 A_0 - \partial_i A_i \equiv \eta_{\mu\nu} \partial^\mu A^\nu, \;\Box = \eta_{\mu\nu} \partial^\mu \partial^\nu \equiv \partial_0^2 - \partial_i^2 $ where the Greek indices 
$\mu, \nu....... = 0, 1, 2, 3$ denote the spacetime directions and the Latin indices $i, j, k.... = 1, 2, 3$ stand for {\it only} the space directions on our flat 4D Minkowskian spacetime manifold.}  
Abelian U(1) gauge theory where there is an explicit coupling between the 
U(1) 1-form gauge field and the charged complex scalar fields. Such a Lagrangian 
density (describing the QED with complex scalar fields) is as follows in the Feynman gauge (see, e.g. [19,23]):  
$$
\begin{array}{lcl}
{\cal L}_B = - {\displaystyle \frac{1}{4} F^{\mu\nu} F_{\mu\nu} + \bar D_\mu \phi^* D^\mu \phi
- V (\phi \phi^*) + B (\partial \cdot A) + \frac{1}{2} B^2 - i \partial_\mu \bar C 
\partial^\mu C},
\end{array}\eqno(2.1)
$$ 
where the covariant derivatives on the fields $\phi$ and $\phi^*$ (with electric charge $e$) are 
$$
\begin{array}{lcl}
D_\mu \phi = \partial_\mu \phi + i e A_\mu \phi,\; \;\qquad \;\;\bar D_\mu \phi^* = 
\partial_\mu \phi^* - i e A_\mu \phi^*.
\end{array}\eqno(2.2)
$$ 
In the above, the U(1) gauge potential $A_\mu$ is derived from the 
1-form connection $A^{(1)} = dx^\mu A_\mu$
which leads to the definition of the curvature 2-form $F^{(2)} = d A^{(1)} \equiv (1/2!) (dx^\mu \wedge dx^\nu) F_{\mu\nu}$ where $d = dx^\mu \partial_\mu$. The curvature tensor $F_{\mu\nu}$ (with $F_{0i} = E_i, F_{ij} = \varepsilon_{ijk} B_k$) is present in the kinetic energy term of the gauge field where
3-vectors $E_i$ and $B_i$ are the electric and magnetic fields and totally antisymmetric
$\varepsilon_{ijk}$ is the 3D Levi-Civita tensor. The Nakanishi-Lautrup auxiliary field $B$ is required to linearize the gauge-fixing term $[- \frac{1}{2} (\partial \cdot A)^2]$ that is present in the Lagrangian density (2.1) in a hidden manner. The fermionic ($C^2 = \bar C^2 = 0,
C \bar C + \bar C C = 0$) (anti-)ghost fields $(\bar C) C$ are required to maintain the unitarity and ``quantum'' gauge (i.e. BRST) invariance {\it together} 
at any arbitrary order of 
the perturbative computations [29]. The gauge (i.e. BRST) invariant potential $V(\phi\phi^*)$ is allowed to include a quartic renormalizable interaction term for the complex scalar fields  
(see, e.g., [30]).

The above Lagrangian density (2.1) is endowed with the following off-shell nilpotent 
($s_{(a)b}^2 = 0$) and anti-commuting  ($s_b s_{ab} + s_{ab} s_b = 0$) (anti-)BRST symmetry transformations $s_{(a)b}$\footnote{We follow here the notations used in [31,32] which imply the fermionic nature of the operators $s_b$ and $s_{ab}$ so that nilpotency property (i.e. $s_{(a)b}^2 = 0$) of these transformations ($s_{(a)b}$) automatically ensue.} for the gauge, (anti-)ghost and complex scalar fields:
$$
\begin{array}{lcl}
&& s_b A_\mu = \partial_\mu C,\; \qquad \; s_b C = 0,\; \qquad s_b \bar C = i B,\; \qquad\; 
s_b \phi = - i e C \phi, \nonumber\\
&& s_b B = 0,\; \qquad s_b F_{\mu\nu} = 0,\; \qquad s_b (\partial \cdot A) = \Box C,\; \qquad s_b \phi^* = + i e \phi^*  C,
\end{array}\eqno(2.3)
$$ 
$$
\begin{array}{lcl}
&& s_{ab} A_\mu = \partial_\mu \bar C,\; \qquad \; s_{ab} \bar C = 0,\; \qquad\; s_{ab} C = - i B,\; \qquad\; s_{ab} \phi = - i e \bar C \phi, \nonumber\\
&& s_{ab} B = 0,\; \qquad s_{ab} F_{\mu\nu} = 0,\; \qquad 
s_{ab} (\partial \cdot A) = \Box \bar C,\;
\qquad s_{ab} \phi^* = + i e \phi^* \bar C.
\end{array}\eqno(2.4)
$$ 
The points to be noted, at this juncture, are as follows:

(i) Under the above nilpotent (anti-)BRST symmetry transformations, the curvature tensor $F_{\mu\nu}$, owing its origin to the exterior derivative 
(i.e. $d = dx^\mu \partial_\mu$) of the differential geometry, remains invariant. This observation is true for the Abelian gauge theory only.

(ii) From the mathematical point of view, the origin of the (anti-)BRST symmetry transformations and their nilpotency property is encoded in the cohomological operator $d = dx^\mu \partial_\mu$ which happens to be a nilpotent operator of order two (i.e. $d^2 = 0$).

(iii) The observation made in (ii) plays a crucial role in the usual superfield approach to BRST formalism [1-12] because its HC is primarily based on the (super) exterior derivatives
(see, e.g. section 3 below for the details of this connection).

(iv) The (anti-)BRST symmetry transformations (2.4) and (2.3), for any arbitrary generic field $\Omega (x)$,  can be succinctly expressed in the mathematical form as
$$
\begin{array}{lcl}
s_r \Omega (x) = - \;i \;[ \Omega,\; Q_r ]_{\pm},\;\;\; \qquad \;\; \;r = b, ab,
\end{array}\eqno(2.5)
$$ 
where $Q_r$ (with $r = b, ab$) are the (anti-)BRST charges that generate the (anti-)BRST transformations in (2.4) and (2.3), $\Omega = A_\mu, C, \bar C, \phi, \phi^*, B$ and $(\pm)$
signs on the square bracket stand for the bracket to be an (anti-)commutator for $\Omega$ being fermionic and bosonic in nature. For our purposes, the explicit form of $Q_r$ is not needed.

(v) The gauge-fixing and Faddeev-Popov terms $[ B (\partial \cdot A) + \frac{1}{2} B^2 - i \partial_\mu \bar C \partial^\mu C]$ of the Lagrangian density (2.1) can be expressed in {\it three} different ways in terms of the off-shell nilpotent (anti-)BRST symmetry transformations $s_{(a)b}$ as given below:
$$
\begin{array}{lcl}
 {\displaystyle s_b \Bigl [- i \;\bar C \{ (\partial \cdot A) + \frac{1}{2} B \} \Bigr ],
\; \quad 
s_{ab} \Bigl [ + i \;C \{ (\partial \cdot A) + \frac{1}{2} B \} \Bigr ],\; \quad
s_b s_{ab} \Bigl [ \frac{i}{2} A_\mu A^\mu + \frac{1}{2} C \bar C \Bigr ]}.
\end{array}\eqno(2.6)
$$
The above expressions are correct modulo some total ordinary spacetime derivative terms.\\

\noindent
{\bf 3 Ramifications of the Horizontality Condition}\\

\noindent
To exploit the potential of the HC in the usual superfield approach to BRST
formalism, first of all, we generalize the ordinary 4D exterior 
derivative $d$ and 1-form connection $A^{(1)}$ to their counterparts on the 
(4, 2)-dimensional supermanifold   
that is parametrized by the usual spacetime variables $x^\mu$ (with $\mu = 0, 1, 2, 3$) 
and a pair of Grassmannian variables $\theta$ and $\bar \theta$ 
(with $\theta^2 = 0, \bar \theta^2 = 0, \theta \bar\theta + \bar\theta \theta = 0$). 
These mappings are (see, e.g. [4,19,23])
 $$
\begin{array}{lcl}
&& d = dx^\mu \partial_\mu\;\;\;\; 
\rightarrow \;\;\;\tilde d = dZ^M \partial_M \equiv 
dx^\mu \partial_\mu + d \theta \partial_\theta + d \bar \theta \partial_{\bar\theta},
\nonumber\\
&& A^{(1)} = dx^\mu A_\mu \;\;\;\rightarrow \;\;\tilde A^{(1)} = d Z^M \tilde A_M \equiv 
dx^\mu {\cal B}_\mu + d \theta \bar {\cal F} + d \bar\theta {\cal F},
\end{array}\eqno(3.1)
$$ 
where $\tilde A_M = ({\cal B}_\mu (x, \theta, \bar\theta,), 
{\cal F} (x, \theta, \bar\theta), \bar {\cal F} (x, \theta, \bar\theta))$ are the 
superfields that are generalization of the ordinary 4D local fields $(A_\mu (x), C (x), \bar C (x))$, $Z^M = (x^\mu,\theta,\bar\theta)$ and $\partial_M = (\partial_\mu, \partial_\theta, \partial_{\bar\theta})$.

The superfields, present in the definition of the super 1-form connection $\tilde A^{(1)}$, can be expanded along the Grassmannian directions of the (4, 2)-dimensional supermanifold  
as 
$$
\begin{array}{lcl}
{\cal B}_\mu  (x, \theta, \bar\theta) &=& A_\mu (x) + \theta\; \bar R_\mu (x) + \bar \theta \;
R_\mu (x) + i \;\theta\; \bar\theta\; S_\mu (x), \nonumber\\
{\cal F} (x, \theta, \bar\theta) &=& C (x) + i \; \theta \; \bar B_1 (x) + i \; \bar \theta\;
B_2 (x) + i \;\theta\; \bar\theta\; s(x), \nonumber\\
\bar {\cal F} (x, \theta, \bar\theta) & = & \bar C (x) + i \; \theta\; \bar B_2 (x) + i \bar \theta\; B_1 (x) + i \;\theta\; \bar\theta\; \bar s (x),
\end{array}\eqno(3.2)
$$ 
where $R_\mu, \bar R_\mu, S_\mu, B_1, \bar B_1, B_2, \bar B_2, s, \bar s$ are the secondary fields that will be determined in terms of the basic fields ($A_\mu, C, \bar C)$ and the Nakanishi-Lautrup auxiliary field $B$ by tapping the potential and power of the HC. It is clear that, in the limit $(\theta, \bar\theta) \to 0$, we retrieve our basic fields from the above expansion (3.2) on the (4, 2)-dimensional supermanifold.

The HC is the mathematical statement that all the Grassmannian components of the super 2-form
curvature $\tilde F^{(2)} = \tilde d \tilde A^{(1)}$ would be set equal to zero so that it becomes equal to the ordinary 2-form curvature $F^{(2)} = d A^{(1)}$. In the language of the physical terms, the above HC is the assertion that the electric and magnetic fields (that are gauge and BRST invariant quantities) should remain independent of the presence of the Grassmannian variables. Thus, the imposition of the restriction $\tilde d \tilde A^{(1)} = d A^{(1)}$ leads to the following relationships [19,23] 
$$
\begin{array}{lcl}
&& R_\mu = \partial_\mu C,\; \qquad \;\bar R_\mu = \partial_\mu \bar C,\; \qquad \;S_\mu = \partial_\mu B, \nonumber\\ && B_2 = \bar B_2 = 0,\; \qquad \;s = \bar s = 0,\; 
\qquad \;B_1 + \bar B_1 = 0.
\end{array}\eqno(3.3)
$$ 
In the above we have the freedom to choose $B_1 = B$ and $\bar B_1 = - B$.

After the application of the HC, we obtain the (anti-)BRST symmetry transformations for the gauge and (anti-)ghost fields {\it together} as is evident from the following expansions
$$
\begin{array}{lcl}
{\cal B}^{(h)}_\mu  (x, \theta, \bar\theta) &=& A_\mu (x) + \theta\;  \partial_\mu \bar C (x) + \bar \theta \;
\partial_\mu  C(x) + i \;\theta\; \bar\theta\; \partial_\mu  B (x) \nonumber\\
&\equiv& A_\mu (x) + \theta \; (s_{ab} A_\mu (x)) + \bar \theta \;(s_b A_\mu (x)) + \theta \; \bar\theta\; (s_b s_{ab} A_\mu (x)), \nonumber\\
{\cal F}^{(h)} (x, \theta, \bar\theta) &=& C (x) - i \; \theta \;  B (x) \nonumber\\ &\equiv& C (x) + \theta \; (s_{ab} C (x)) + \bar\theta \; (s_b C (x)) + \theta\; \bar\theta\; (s_b s_{ab} C (x)), \nonumber\\
\bar {\cal F}^{(h)} (x, \theta, \bar\theta) & = & \bar C (x) + i \;\bar \theta\; B_(x) \nonumber\\ &\equiv& \bar C (x) + \theta \; (s_{ab} \bar C (x)) + \bar \theta (s_b \bar C (x)) + \theta\; \bar\theta\; (s_b s_{ab} \bar C (x)),
\end{array}\eqno(3.4)
$$ 
where we have taken into account the fact that $s_b C (x) = 0$ and $s_{ab} \bar C (x) = 0$. The above expansions clearly imply that $\tilde F^{(2)}_{(h)} = F^{(2)}$ where $\tilde F^{(2)}_{(h)} = \tilde d \tilde A^{(1)}_{(h)}$ is the super curvature 2-form expressed in terms of the superfields in (3.4).

The main consequences of the application of the HC are:

(i) The geometrical interpretations of the (anti-)BRST symmetry transformations and their corresponding generators as the translational generators along the Grassmannian directions of the (4, 2)-dimensional supermanifold (cf. (2.5) and (3.4))
$$
\begin{array}{lcl}
{\displaystyle s_b \Leftrightarrow Q_b \Leftrightarrow \mbox{Lim}_{\theta \to 0} \frac{\partial}{\partial \bar\theta},\; \qquad\;
s_{ab} \Leftrightarrow Q_{ab} \Leftrightarrow \mbox{Lim}_{\bar \theta \to 0} \frac{\partial}{\partial \theta}}.
\end{array}\eqno(3.5)
$$ 
The above mapping also sheds light on the nilpotency and anticommutativity properties of the 
(anti-)BRST symmetry transformations and their corresponding generators [19,23].

(ii) The kinetic energy term $- \frac{1}{4} F^{\mu\nu} F_{\mu\nu}$ of the U(1) Abelian 1-form gauge field remains intact
when it is expressed in terms of the superfield ${\cal B}_\mu^{(h)}$ because
$$
\begin{array}{lcl}
\tilde F_{\mu\nu}^{(h)} = \partial_\mu {\cal B}_\nu^{(h)} - \partial_\nu {\cal B}_\mu^{(h)}
\equiv F_{\mu\nu} \;\;\;\Rightarrow \;\;\;{\displaystyle - \frac{1}{4} \tilde F^{\mu\nu (h)} \tilde F_{\mu\nu}^{(h)} = - \frac{1}{4} F^{\mu\nu} F_{\mu\nu}}.
\end{array}\eqno(3.6)
$$ 
In other words, expressed in terms of the superfields, the kinetic energy term (cf. (2.1)) of the gauge field remains independent of the Grassmannian variables $\theta$ and $\bar\theta$.

(iii) As a consequence of the mappings in (3.5), we can express the (anti-)BRST exact expressions, illustrated in (2.6), as follows 
$$
\begin{array}{lcl}
&& \mbox{Lim}_{\theta \to 0} {\displaystyle \frac{\partial}{\partial \bar\theta} \Bigl [
- i\;\bar {\cal F}^{(h)} \bigl \{ (\partial \cdot {\cal B}^{(h)} + \frac{1}{2} B \bigr \} \Bigr ]}, \quad
 \mbox{Lim}_{\bar \theta \to 0} {\displaystyle \frac{\partial}{\partial \theta} \Bigl [
+ i \;  {\cal F}^{(h)} \bigl \{ (\partial \cdot {\cal B}^{(h)} + \frac{1}{2} B \bigr \} \Bigr ]},
\nonumber\\
&& {\displaystyle \frac{\partial}{\partial \bar\theta} \; \frac{\partial}{\partial\theta}
\Bigl [ \frac{i}{2} {\cal B}^{(h)} \cdot {\cal B}^{(h)} + \frac{1}{2} {\cal F}^{(h)} 
\bar {\cal F}^{(h)} \Bigr ]}. 
\end{array}\eqno(3.7)
$$

(iv) The (anti-)BRST invariance of the kinetic energy term of the U(1) gauge field as well as the sum of the gauge-fixing and Faddeev-Popov ghost terms of the 4D Lagrangian density (2.1) can be captured in the language of the superfield approach to BRST formalism. To this end in mind, first of all, we note that the corresponding super Lagrangian density $\tilde {\cal L}^{(1)}$
will be the sum of (3.6) and (3.7). Mathematically, this statement can be expressed, in three different ways, as listed below
$$
\begin{array}{lcl}
\tilde {\cal L}^{(1)}_{(i)} = {\displaystyle - \frac{1}{4} \tilde F^{\mu\nu (h)} \tilde F_{\mu\nu}^{(h)}} +  \mbox{Lim}_{\theta \to 0} {\displaystyle \frac{\partial}{\partial \bar\theta} \;\Bigl [\;
- i\;\bar {\cal F}^{(h)} \bigl \{ (\partial \cdot {\cal B}^{(h)} + \frac{1}{2} B \bigr \} 
\;\Bigr ]},
\end{array}\eqno(3.8)
$$ 
$$
\begin{array}{lcl}
\tilde {\cal L}^{(1)}_{(ii)} = {\displaystyle - \frac{1}{4} \tilde F^{\mu\nu (h)} \tilde F_{\mu\nu}^{(h)}} +  \mbox{Lim}_{\bar \theta \to 0} {\displaystyle \frac{\partial}{\partial \theta} \;\Bigl [\;
+ i\; {\cal F}^{(h)} \bigl \{ (\partial \cdot {\cal B}^{(h)} + \frac{1}{2} B \bigr \} \;\Bigr ]},
\end{array}\eqno(3.9)
$$ 
$$
\begin{array}{lcl}
\tilde {\cal L}^{(1)}_{(iii)} = {\displaystyle - \frac{1}{4} \tilde F^{\mu\nu (h)} \tilde F_{\mu\nu}^{(h)}} + {\displaystyle \frac{\partial}{\partial \bar\theta} \; \frac{\partial}{\partial\theta}\;
\Bigl [ \;\frac{i}{2} {\cal B}^{(h)} \cdot {\cal B}^{(h)} + \frac{1}{2} {\cal F}^{(h)} 
\bar {\cal F}^{(h)}\; \Bigr ]}. 
\end{array}\eqno(3.10)
$$
The BRST and anti-BRST invariance of the corresponding 4D Lagrangian density
$$
\begin{array}{lcl}
s_{(a)b} {\cal L}^{(1)} = 0 \quad \mbox{with} \quad {\cal L}^{(1)} = - {\displaystyle 
\frac{1}{4} F^{\mu\nu} F_{\mu\nu} + B (\partial \cdot A) + \frac{B^2}{2}
- i \partial_\mu \bar C \partial^\mu C},
\end{array}\eqno(3.11)
$$ 
can be captured in the language of the superfield formulation as 
$$
\begin{array}{lcl}
&& \mbox{Lim}_{\theta \to 0} {\displaystyle \frac{\partial}{\partial \bar \theta}
\; \tilde {\cal L}^{(1)}_{(i)} = 0 \;\;\;\Leftrightarrow \;\;\; s_b {\cal L}^{(1)} = 0},
\nonumber\\
&& \mbox{Lim}_{\bar \theta \to 0} {\displaystyle \frac{\partial}{\partial \theta}}
\; \tilde {\cal L}^{(1)}_{(ii)} = 0 \;\;\;\Leftrightarrow \;\;\; s_{ab} {\cal L}^{(1)} = 0, 
\nonumber\\
&& \Bigl ( {\displaystyle \frac{\partial}{\partial \theta} \; \tilde {\cal L}^{(1)}_{(iii)} = 0, \;
\frac{\partial}{\partial \bar\theta}\; \tilde {\cal L}^{(1)}_{(iii)} = 0} \Bigr )\; \Leftrightarrow 
s_{(a)b} {\cal L}^{(1)} = 0.
\end{array}\eqno(3.12)
$$ 
In other words, the Grassmannian independence of the super Lagrangian densities (cf. (3.6) and (3.8)--(3.10)) imply the (anti-)BRST invariance of the kinetic energy term for the U(1) gauge field as well as the sum of the gauge-fixing and Faddeev-Popov ghost terms of the 4D Lagrangian density ${\cal L}^{(1)}$ (cf. (3.11)).\\

\noindent
{\bf 4 Consequence of the Gauge Invariant Restriction on the Matter Superfields }\\

\noindent
Let us begin with the gauge-invariant restriction (GIR) on the matter superfields
$\Phi (x, \theta, \bar\theta)$ and $\Phi^* (x, \theta, \bar\theta)$ where the HC of the previous section plays a very decisive role. This crucial
and cute restriction is as follows [19]
$$
\begin{array}{lcl}
\Phi^* (x, \theta, \bar\theta) \;\bigl (\tilde d + i e \tilde A^{(1)}_{(h)} \bigr)\;
\Phi (x, \theta, \bar\theta) = \phi^* (x)\; \bigl (d + i e A^{(1)} \bigr ) \; \phi (x),
\end{array}\eqno(4.1)
$$
where $\tilde A^{(1)}_{(h)} = dx^\mu {\cal B}_\mu^{(h)} + d \theta \bar {\cal F}^{(h)}
+ d \bar\theta {\cal F}^{(h)}$ is the super 1-form connection obtained after the application of the HC\footnote{It is self-evident that all the quantities on the l.h.s. of (4.1) are defined on the (4, 2)-dimensional supermanifold which is paramaterized by the superspace variable $Z^M = (x^\mu, \theta, \bar\theta)$.}. It is elementary to check that the r.h.s. of the above equation (i.e. 
$dx^\mu \phi^* (x) D_\mu \phi (x)$) is a gauge (i.e. BRST) invariant quantity. The noteworthy point is the fact that there is an interplay between the HC and the GIR in relationship (4.1)
where the 1-form gauge field and the matter fields are intertwined {\it together} in a useful manner.

In the above equation, all the symbols carry their standard meaning and the super  matter fields $\Phi (x, \theta, \bar\theta)$ and $\Phi^* (x, \theta, \bar\theta)$ are the generalization of the 4D local matter fields $\phi(x)$ and $\phi^* (x)$  on the (4, 2)-dimensional supermanifold. The expansion of the above superfields along the Grassmannian directions of the (4, 2)-dimensional supermanifold, are 
$$
\begin{array}{lcl}
\Phi (x,\theta,\bar\theta) &=& \phi (x) + i \;\theta\; \bar f_1 (x) + i \;\bar\theta\; f_2 + i \;\theta \; \bar\theta\; b (x), \nonumber\\
\Phi^* (x,\theta,\bar\theta) &=& \phi^* (x) + i \;\theta\; \bar f^*_2 (x) + i \;\bar\theta\; f^*_1 + i\; \theta \; \bar\theta\; \bar b^* (x),
\end{array}\eqno(4.2)
$$
where the secondary fields ($f^*_1, \bar f_1, f_2, \bar f^*_2, b, \bar b^*$) would be
determined from the GIR (4.1). It is evident that, in the limit $(\theta, \bar\theta) \to 0$,
the above superfields reduce to the 4D local fields $\phi (x)$ and $\phi^* (x)$. In fact, in our earlier work, we have computed 
the exact expressions of the secondary fields, in terms of the basic and auxiliary fields, as [19]
$$
\begin{array}{lcl}
&& \bar f_1 = - e \bar C \phi,\; \qquad f_2 = - e C \phi,\; \qquad 
b = - i \;e \;(B + e \bar C C) \; \phi,
\nonumber\\
&& f_1^* = e C \phi^*, \; \qquad \;\bar f_2^* = e \bar C \phi^*,\; 
\qquad \;\bar b^* = i \;e \;(B + e C \bar C) \;\phi^*.
\end{array}\eqno(4.3)
$$
In the above explicit computation, the power of the GIR (cf. (4.1)) and the expansion obtained after the application of the HC (cf. (3.4)) have been exploited
together.

With inputs from (4.3), we have the following expansions for the super matter fields 
$$
\begin{array}{lcl}
\Phi_{(g)} (x,\theta,\bar\theta) &=& \phi  + \theta\; (- i e \bar C) \phi +  \bar\theta\; 
(- i e C) \phi + e\;\theta \; \bar\theta\; (B + e \bar C C) \phi \nonumber\\
&\equiv& \phi  + \theta\; (s_{ab} \phi) + \bar\theta\; (s_b \phi) + \theta\;\bar\theta\;
(s_b s_{ab} \phi), \nonumber\\
\Phi^*_{(g)} (x,\theta,\bar\theta) &=& \phi^* (x) + \theta\; (i e \bar C) \phi^*  +  \bar\theta\;(i e C) \phi^*  -e\; \theta \; \bar\theta\; ( B + e C \bar C) \phi^*, \nonumber\\
&\equiv& \phi^* +  \theta\; (s_{ab} \phi^*) + \bar\theta\; (s_b \phi^*) + \theta\;\bar\theta\;
(s_b s_{ab} \phi^*),
\end{array}\eqno(4.4)
$$
which, very naturally, lead to the derivation of the precise expressions for the (anti-)BRST symmetry transformations for the matter fields (cf. (2.3), (2.4)). Furthermore, the above expansions provide
the same useful mappings as listed in (3.5). In other words, the GIR in (4.1) is consistent with the HC of the usual superfield approach to the BRST formalism because the geometrical interpretations (and origin) of the (anti-)BRST symmetry transformations (and their corresponding generators) remain the same.

It is now very interesting to know the exact form of a part 
(i.e. ${\cal L}^{(2)} = \bar D_\mu \phi^* D^\mu \phi - V (\phi \phi^*)$) of the 4D Lagrangian density (2.1) in the language of the superfields defined on the  (4, 2)-dimensional supermanifold. In this connection, it is worthwhile to point out that the super expansion (4.4), obtained after the application of the GIR (4.1), is such that the following equalities are automatically satisfied:
$$
\begin{array}{lcl}
\Phi^*_{(g)} (x, \theta, \bar\theta)\; \Phi_{(g)} (x, \theta, \bar\theta) = \phi^* (x)\; 
\phi (x) \;\;\;\Rightarrow\;\;\;\;\; \tilde V (\Phi^*_{(g)} \Phi_{(g)}) = V (\phi^*\phi). 
\end{array}\eqno(4.5)
$$
Stated in a different way, the form of the gauge (i.e. BRST) invariant potential
 $V(\phi^*\phi)$ remains independent of the Grassmannian variables when its super-version
(i.e. $\tilde V (\Phi^*_{(g)} \Phi_{(g)})$) is expressed in terms of the expansions of the matter superfields  in (4.4).

With only the GIR on the matter superfields at our disposal (cf. (4.1)), it is {\it not} clear how to express the kinetic energy term
(i.e. $\bar D_\mu \phi^* D^\mu \phi$) of the 4D local matter fields in the language of the superfields defined on the (4, 2)-dimensional supermanifold. However, it is very interesting 
to check that the following equality\footnote{It is evident that the r.h.s. of the equality in (4.6) is also a gauge (i.e. BRST) invariant quantity. Instead of (4.1), we could have started from (4.6) itself as a GIR on the (4, 2)-dimensional supermanifold to derive the relationship in (4.1). However, this way of calculating things would be very cumbersome 
and complicated because the results of (4.3) would not emerge very easily.}
$$
\begin{array}{lcl}
\tilde {\bar D}^{(h)}_M \Phi^*_{(g)} (x,\theta,\bar\theta) \; \tilde D^{M (h)} \Phi_{(g)} (x, \theta, \bar\theta) = \bar D_\mu \phi^* (x)\; D^\mu \phi (x),
\end{array}\eqno(4.6)
$$
holds good where the l.h.s. is explicitly expressed as follows 
$$
\begin{array}{lcl}
&& \tilde {\bar D}^{(h)}_M \Phi^*_{(g)} (x,\theta,\bar\theta) \; \tilde D^{M (h)} \Phi_{(g)} (x, \theta, \bar\theta) = (\partial_\mu - i e {\cal B}^{(h)}_\mu) \Phi^*_{(g)}\; 
(\partial^\mu + i e {\cal B}^{\mu (h)}) \Phi_{(g)}, \nonumber\\
&& + (\partial_\theta - i e \bar {\cal F}^{(h)}) \Phi^*_{(g)} \;(\partial_\theta + i e \bar {\cal F}^{(h)}) \Phi_{(g)} 
+ (\partial_{\bar\theta} - i e  {\cal F}^{(h)}) \Phi^*_{(g)} (\partial_{\bar\theta} + i e  {\cal F}^{(h)}) \Phi_{(g)}.
\end{array}\eqno(4.7)
$$
A mere look at the above equation does not imply that the equality in (4.6) is precisely valid. However,
it is very interesting to point out that the following equations 
$$
\begin{array}{lcl}
(\partial_\theta + i e \bar {\cal F}^{(h)}) \;\Phi_{(g)} = 0,\;\; \qquad\;\; 
(\partial_{\bar\theta} + i e  {\cal F}^{(h)}) \;\Phi_{(g)} = 0,
\end{array}\eqno(4.8)
$$
are automatically satisfied due to the expansions in (3.4) and (4.4). As a consequence, the {\it second and third} terms of the r.h.s. of the expansion in (4.7) turn out to be {\it zero}.

It is gratifying, at this stage, to point out that the remaining first term of the r.h.s of the expansion in (4.7) leads to the following equality
$$
\begin{array}{lcl}
(\partial_\mu - i e {\cal B}^{(h)}_\mu)\; \Phi^*_{(g)}\; 
(\partial^\mu + i e {\cal B}^{\mu (h)})\; \Phi_{(g)} = (\partial_\mu - i e A_\mu)\; \phi^* (x)\; 
(\partial^\mu + i e A^{\mu})\; \phi (x).
\end{array}\eqno(4.9)
$$
In other words, the expansion (3.4) (obtained after the application of the HC) and the super expansion of (4.4) (deduced after the application of the HC as well as GIR in (4.1))
are such that the equality listed in (4.6) is sacrosanct.

A key consequence of the equality in (4.6) is the fact that the l.h.s., even though expressed in terms of the (4, 2)-dimensional superfields, is independent of the Grassmannian variables 
$\theta$ and $\bar\theta$. As a result, the r.h.s. of the equation (4.6) (which is nothing but the kinetic energy term for the complex scalar fields) turns out to be the (anti-)BRST invariant (i.e. $s_{(a)b} [\bar D_\mu \phi^* D^\mu \phi] = 0$) quantity. This can be checked explicitly by exploiting the definitions (2.2) and transformations (2.3) as well as (2.4).
Ultimately, we obtain the following expression of the super Lagrangian density $\tilde {\cal L}^{(2)}$: 
$$
\begin{array}{lcl}
\tilde {\cal L}^{(2)} = \tilde {\bar D}^{(h)}_M \Phi^*_{(g)} (x,\theta,\bar\theta) \; \tilde D^{M (h)} \Phi_{(g)} (x, \theta, \bar\theta) - \tilde V (\Phi^*_{(g)} \Phi_{(g)}),
\end{array}\eqno(4.10)
$$
which is nothing but the generalization of the 4D Lagrangian density ${\cal L}^{(2)} = \bar D_\mu \phi^* D^\mu \phi - V (\phi \phi^*)$ to the (4, 2)-dimensional supermanifold.

The (anti-)BRST invariance of the 4D Lagrangian density ${\cal L}^{(2)}$ can be expressed, in the language of the augmented superfield approach to BRST formalism, as:
$$
\begin{array}{lcl}
&& \mbox{Lim}_{\theta \to 0} {\displaystyle \frac{\partial}{\partial \bar \theta}
\; \tilde {\cal L}^{(2)}  = 0 \;\;\;\Leftrightarrow \;\;\; s_b {\cal L}^{(2)} = 0}, 
\nonumber\\
&& \mbox{Lim}_{\bar \theta \to 0} {\displaystyle \frac{\partial}{\partial \theta}}
\; \tilde {\cal L}^{(2)} = 0 \;\;\;\Leftrightarrow \;\;\; s_{ab} {\cal L}^{(2)} = 0, \nonumber\\
&& \Bigl ( {\displaystyle \frac{\partial}{\partial \theta}}
\; \tilde {\cal L}^{(2)}  = 0, {\displaystyle \frac{\partial}{\partial \bar \theta}}
\; \tilde {\cal L}^{(2)} = 0 \Bigr ) \;\;\;\Leftrightarrow\;\;\; s_{(a)b} {\cal L}^{(2)} = 0.  
\end{array}\eqno(4.11)
$$
Thus, we note that the (anti-)BRST invariance of the total 4D Lagrangian density (2.1) of the QED with charged complex scalar fields is concisely captured in the language of the superfield approach to BRST formalism by the equations (3.12) and (4.11).\\

\noindent
{\bf 5 Impact of the Single Gauge Invariant Restriction on the Matter Superfields}\\

\noindent
In our very recent works [23-25], we have been able to generalize the HC in such a way that we 
can deduce all the results of our sections 3 and 4 in a {\it single} stroke by imposing a GIR on the matter superfields. In this condition, a pair of (super) covariant derivatives and their intimate connection with the (super) 2-form curvatures $(\tilde F^{(2)}) F^{(2)}$ are exploited. These relationships can be expressed  in two different ways as
[24,25] 
$$
\begin{array}{lcl}
&&\Phi^* (x, \theta, \bar\theta) \;\tilde D \;\tilde D \;\Phi (x,\theta,\bar\theta) = \phi^* (x) \;D \;D\; \phi (x), \nonumber\\
&& \Phi (x, \theta, \bar\theta) \;\tilde {\bar  D}\; \tilde {\bar D} \;\Phi^* (x,\theta,\bar\theta) = \phi (x) \;\bar D \;\bar D \;\phi^* (x),
\end{array}\eqno(5.1)
$$
where the (super) covariant derivatives have the following explicit forms:
$$
\begin{array}{lcl}
\tilde D = \tilde d + i e \tilde A^{(1)},\; \quad\;
\tilde {\bar D} = \tilde d - i e \tilde A^{(1)},\; \quad\; D = d + i e A^{(1)},\; \quad \;
\bar D = d - i e A^{(1)}.
\end{array}\eqno(5.2 )
$$
The symbols, used in the above definitions, have been explained in our earlier sections.

Expressed in terms of the superfields and super differentials, we have 
$$
\begin{array}{lcl}
&&\tilde D = dx^\mu \;(\partial_\mu + i e {\cal B}_\mu) + d \theta \;(\partial_\theta + i e 
\bar {\cal F}) + d \bar\theta \;(\partial_{\bar\theta} + i e {\cal F}), \nonumber\\
&&\tilde {\bar D} = dx^\mu \;(\partial_\mu - i e {\cal B}_\mu) + d \theta \;(\partial_\theta - i e 
\bar {\cal F}) + d \bar\theta \;(\partial_{\bar\theta} - i e {\cal F}).
\end{array}\eqno(5.3)
$$
It is straightforward to check that the r.h.s. of both the expressions of (5.1) are gauge (i.e. BRST) invariant quantities because they can be explicitly expressed as 
$$
\begin{array}{lcl}
i \;e \;\phi^* (x) \;F^{(2)} \;\phi (x),\;\;\; \qquad \;\;\;
- i \;e \;\phi (x) \;F^{(2)} \;\phi^* (x).
\end{array}\eqno(5.4)
$$
It is elementary to state that the above quantities are invariant under the U(1) gauge transformations: $\phi (x) \to U\; \phi(x), F^{(2)} \to F^{(2)}, \phi^* (x) \to \phi^* (x)\; U^{-1} \equiv U^{-1}\; \phi^* (x)$ where $U$ is an arbitrary exponential transformation corresponding to the Abelian U(1) gauge group.

The salient features of the GIR in (5.1) are as follows. First and foremost, all the results, quoted in (3.3) and (4.3), are obtained from this {\it single} restriction (see, e.g. [24,25] for details). Thus, the condition (5.1) is theoretically more economical than the HC of section 3 and GIR of section 4 which are exploited separately and independently. Second, it allows the derivation of all the (anti-)BRST symmetry transformations for {\it all} the fields of the theory {\it together}. Finally, the basic ideas of the gauge theories blend together in a beautiful manner in (5.1) because of the fact that the matter fields, the 1-form gauge connections, the covariant derivatives and their intimate relations with the curvature 2-forms, etc., are found to be beautifully 
intertwined together, at one place, in (5.1).

Since the outcome of the condition (5.1) is same as our results of sections 3 and 4, it is
straightforward to express the total super Lagrangian density of the theory as 
$$
\begin{array}{lcl}
\tilde {\cal L}_T\; = \;\tilde {\cal L}^{(1)} \;+ \; \tilde {\cal L}^{(2)},
\end{array}\eqno(5.5)
$$
where $\tilde {\cal L}^{(1)}$ is expressed in three different forms in (3.8)--(3.10) and 
$\tilde {\cal L}^{(2)}$ is expressed in (4.10). The (anti-)BRST invariance of the 4D Lagrangian density (2.1) can be expressed, in the language of the the superfield approach to BRST formalism, as 
$$
\begin{array}{lcl}
&& \mbox{Lim}_{\theta \to 0} {\displaystyle \frac{\partial}{\partial \bar \theta}}
\; \tilde {\cal L}_T  = 0 \;\;\;\Leftrightarrow \;\;\; s_b {\cal L}_B = 0, 
\nonumber\\
&& \mbox{Lim}_{\bar \theta \to 0} {\displaystyle \frac{\partial}{\partial \theta}}
\; \tilde {\cal L}_T = 0 \;\;\;\Leftrightarrow \;\;\; s_{ab} {\cal L}_B = 0, \nonumber\\
&& \Bigl ( {\displaystyle \frac{\partial}{\partial \bar \theta}}
\; \tilde {\cal L}_T  = 0, {\displaystyle \frac{\partial}{\partial \theta}}
\; \tilde {\cal L}_T = 0 \Bigr ) \;\;\;\Leftrightarrow\;\;\; s_{(a)b} {\cal L}_B = 0. 
\end{array}\eqno(5.6)
$$
Finally, we note that there is a great deal of simplification in the understanding of the (anti-)BRST invariance of the 4D QED with charged complex scalar fields if we exploit the theoretical arsenal of the augmented superfield approach to BRST formalism.\\

\noindent {\bf 6 Conclusions}\\

\noindent
One of the central objectives of our present investigation was to understand the geometrical meaning of the (anti-)BRST invariance, present in the context of the 4D QED with charged complex scalar  fields, in the terminology of the augmented superfield approach to BRST formalism [13-25].
We have accomplished that goal in equation (5.6) where we have been able to demonstrate that
if the operation of the partial derivatives w.r.t. the Grassmannian variables on the (4, 2)-dimensional super Lagrangian density turns out to be zero, the corresponding 4D Lagrangian density would certainly respect (anti-)BRST invariance.

Geometrically, the essence of equation (5.6) implies that if the super Lagrangian density of the theory is translated along the Grassmannian $\bar\theta$-direction of the (4, 2)-dimensional supermanifold\footnote{When there is {\it no} translation along the Grassmannian $\theta$-direction of the supermanifold (i.e. $\theta \to 0$).} such that the outcome of this shift operation is zero, the corresponding 4D Lagrangian density is positively endowed with the nilpotent BRST symmetry invariance. If the above operation is true when the Grassmannian variables ($\theta$ and $\bar\theta$) are exchanged 
with each-other (i.e. $\theta \leftrightarrow \bar\theta$) , there
is presence of the nilpotent anti-BRST symmetry invariance because the corresponding 4D Lagrangian density will remain unchanged (or quasi-invariant) under the nilpotent anti-BRST symmetry transformations.

One of the key observations in our present investigation is that the equality in (4.6) is true for the GIR imposed in (4.1) on the matter superfields. In other words, the GIR in (4.1) 
implies sanctity of the equality in (4.6) which enables us, in turn, to express the 4D kinetic energy term 
(i.e. $\bar D_\mu \phi^* D^\mu \phi$) for
the ordinary 4D local matter fields (i.e. $\phi (x)$ and $\phi^* (x)$) in the language of the (4, 2)-dimensional matter superfields that are derived in (4.4) after
the application of GIR. Furthermore, the expansions in (4.4) also play a central role in expressing the potential energy term (i.e. $V(\phi \phi^*)$) of the charged complex scalar fields in the language of the (4, 2)-dimensional superfields (cf. (4.10)).

It should be noted that the equality in (3.6) and (4.6) immediately imply that the kinetic energy terms of the $U(1)$ gauge field and charged complex scalar fields 
are invariant under the (anti-)BRST symmetry transformations (2.3) and (2.4). Similarly, the nilpotent (anti-)BRST symmetry invariance of the potential energy term of the matter fields $\phi$ and $\phi^*$ (i.e. $V (\phi\phi^*)$) 
is encoded in (4.5). The expressions for the gauge-fixing and Faddeev-Popov ghost terms in (3.7) lead to the conclusion that these terms are also invariant under the (anti-)BRST symmetry transformations
(cf. (2.3) and (2.4)) due to the nilpotency 
(i.e. $\partial_\theta^2 = \partial_{\bar\theta}^2 = 0$) 
of the translational generators $\partial_\theta$ and $\partial_{\bar\theta}$.

The main result of our present paper should be taken to be our attempt in providing a complete geometrical picture of the (anti-)BRST symmetry transformations, their nilpotency and anticommutativity properties and their invariance within the framework of the superfield approach to BRST formalism. At least, for the (non-)Abelian 1-form gauge theories,
we have been able to furnish the geometrical interpretations of all the above mentioned
properties within the framework of the superfield approach to BRST formalism.

One of us has been involved with a different kind of superfield approach to BRST formalism
[33-39] which has been christened as the BRST superspace formulation. This formulation has also been applied to study the 1-form gauge theories where the Ward-Takahashi identities emerge very naturally. Furthermore, one has almost complete freedom to discuss
the translations, rotations, dilatations, etc., of the Grassmannian variables. This formulation has a generalized gauge invariance in the superspace and is very suitable in the study of the renormalization of the gauge invariant operators and the gauge theories itself.

It would be nice endeavor to unify the above two types of the superfield approaches to BRST formalism and apply it to some gauge theories. Furthermore, it would be nice to apply the geometrical superfield approach to 2-form gauge theories which are very important in the context of string theories. A couple of modest attempts, in this direction, have already been made in [39,40]
for the case of 4D free Abelian 2-form gauge theory. In fact, it is because of the latter study [40] that we have been able to derive an off-shell nilpotent and {\it absolutely} anticommuting (anti-)BRST symmetry transformations for the above theory [41]. This result has a deep connection with the concept of gerbes.

It would be very fruitful venture to study the 4D non-Abelian 2-form gauge theories in the framework of the superfield approach to BRST formalism and see its connection with the non-Abelian gerbes. We, furthermore, plan to apply the superfield approach to 4D gravitational theories where an attempt [7] has already been made. These are some of the promising issues that are presently under investigation and our results would be reported in our forthcoming future publications.\\

\noindent {\bf Acknowledgements}\\

\noindent
One of us (RPM) would like to gratefully acknowledge the financial support from the
Department of Science and Technology (DST), Government of India,  
under the SERC project sanction grant No: - SR/S2/HEP-23/2006.

\end{document}